\documentclass[twocolumn,twoside,slac_two]{revtex4}
\usepackage{graphicx} 
\usepackage{fancyhdr}
\usepackage{hyperref}
\hypersetup{
    colorlinks=true,
    urlcolor=magenta,
}
\begin{document}

\title{Space Radiation exposure calculations during different solar and galactic cosmic ray activities}

%

\author{P. Paschalis}
\affiliation{Nuclear and Particle Physics Department. Faculty of Physics, National and Kapodistrian University of Athens, 15784 Athens Greece}
\author{A. Tezari}
\affiliation{Nuclear and Particle Physics Department. Faculty of Physics, National and Kapodistrian University of Athens, 15784 Athens Greece \& Medical School, National and Kapodistrian University of Athens, 11527 Athens Greece}
\author{M. Gerontidou, H. Mavromichalaki}
\affiliation{Nuclear and Particle Physics Department. Faculty of Physics, National and Kapodistrian University of Athens, 15784 Athens Greece.}
\author{P. Nikolopoulou}
\affiliation{Medical School, National and Kapodistrian University of Athens, 11527 Athens Greece}

\begin{abstract}
The primary components of radiation in interplanetary space are galactic cosmic rays (GCR) and solar cosmic radiation (SCR). GCR originates from outside of our Solar System, while SCR consists of low energy solar wind particles that flow constantly from the Sun and the highly energetic solar particle events (SPEs) that originate from magnetically disturbed regions of the Sun, which sporadically emit bursts of energetic charged particles. Exposure to space radiation may place astronauts and aviation crews at significant risk for numerous biological effects resulting from exposure to radiation from a major SPE or combined SPE and GCR. Doses absorbed by tissues vary for different SPEs and model systems have been developed to calculate the radiation doses that could have been received by astronauts during previous SPEs. For this reason a new application DYASTIMA-R which constitutes a successor of the Dynamic Atmospheric Shower Tracking Interactive Model Application (DYASTIMA) is being developed. This new simulation tool will be used for the calculation of the equivalent dose during flights scenario in the lower or higher atmosphere, characterized by different altitudes, different geographic latitudes and different solar and galactic cosmic ray intensity. Results for the calculated energy deposition and equivalent dose are calculated during quiet and disturbed periods of the solar cycles 23 and 24, are presented.
\end{abstract}

\maketitle

\thispagestyle{fancy}

\section{INTRODUCTION}
The general systems of radiation protection on Earth are not appropriate for the study of radiation exposure of aviation crews\cite{cite1}, since high energy charged particles contribute significantly to the total dose in the human body \cite{cite2}. These are due to the interaction of the Galactic Cosmic Rays (GCR) and solar energetic particles (SEP) with the atmospheric layers and the cascades of the secondary particles that are produced \cite{cite3}. As the altitude in-creases, the atmospheric protective layer gets thinner and less dense, resulting in higher cosmic radiation (CR) than the radiation on the ground.

Radiation in the different altitudes of the atmosphere is ionizing (GCR, SEP and trapped radiation inside Earth’s magnetic field) and non-ionizing (UV-radiation), with many biological and technological effects. As far as human health and exposure is concerned the effects can be acute (nausea / vomiting, fatigue, central nervous system disease) and chronic (cancer / solid tumors / leukemia, cataract / vision impairment, degenerative cardiac disease) \cite{cite1}. Ionizing radiation is by far more dangerous and the acute effects after exposure are related to the high intensity SPE, while the chronic effects are due to long term exposure to GCR \cite{cite2}. For example, the mean equivalent dose during a 7-hrs flight is 0.05 mSv for a quiet period, while for an extreme SPE (105 particles / cm2 st sec) it raises up to 40 mSv. It is noted that the average equivalent dose is 1 mSv / year for public exposure according to the European Directive 2013/59/ Euratom. Therefore, one of the primary concerns for aircraft flights is the elevated level of radiation that aviators and passengers are exposed.

Several models are developed in order to study the atmospheric showers such as the CRII model\cite{cite4}, PLANETO-COSMICS\cite{cite5} and DYASTIMA \cite{cite3}. At the same time a number of applications are used for the determination of biological and technological effects and the calculation of the radiation exposure such as SPENVIS-CREME \cite{cite6}, SIEVERT\cite{cite7}, NAIRAS\cite{cite8} and AVIDOS\cite{cite9}. A new application, named DYASTIMA-R, which calculates the equivalent dose in different altitudes during quiet and disturbed solar activity periods is being developed and is presented in this work.

\section{DYASTIMA}
In order to implement a simulation of the cosmic ray propagation through the atmosphere, there are some physical quantities and processes that must be taken into consideration, such as the spectrum of the primary CR that reach the top of the atmosphere, the structure of the atmosphere, the Earth's magnetic field and the physical interactions that take place between the CR particles and the molecules of the atmosphere. These quantities are affected by various parameters, such as the space weather conditions, the current physical characteristics of the Earth's atmosphere, the time and the location for which the simulation is performed \cite{cite3}. 

DYASTIMA is a standalone application for the simulation of the showers that are produced in the atmosphere of a planet due to the CR. The application makes use of the well known Geant4 simulation toolkit \cite{cite3} \cite{cite10} \cite{cite11}. The simulation scenario is described by using a graphical user interface (GUI) and requires as input all the parameters mentioned above. The output of DYASTIMA provides all the available information about the cascade and tracking, such as number, energy, direction, arrival time and energy deposit of the secondary particles at different atmospheric layers. DYASTIMA is also used for cascades simulation in the atmosphere of other planets \cite{cite12}.

\section{DYASTIMA-R}
\begin{figure}[h]
\centering
\includegraphics[width=0.6\textwidth,bb=0 0 420 350]{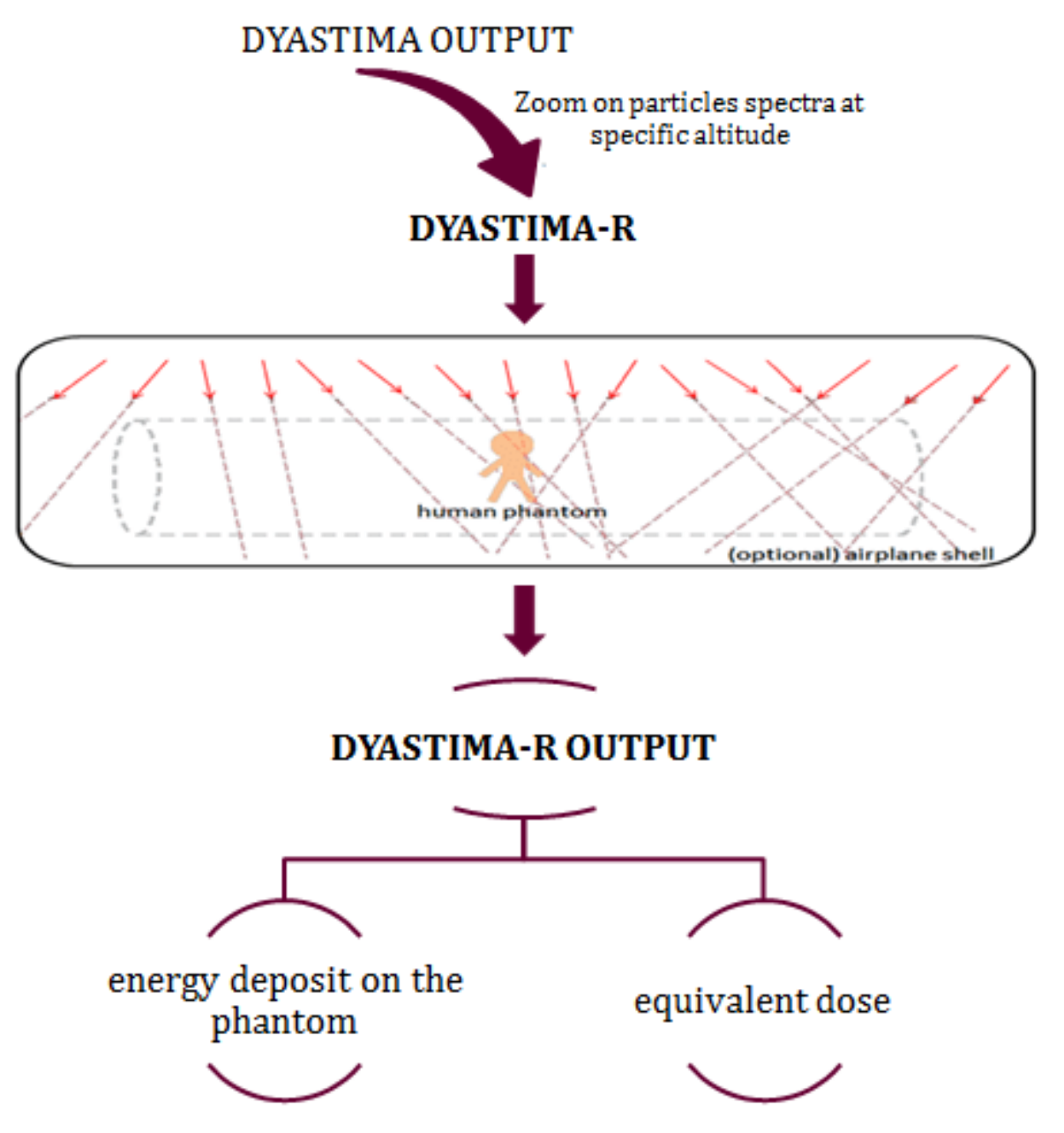}
\vspace*{-0.7cm}
\caption{Block diagram for DYASTIMA-R.}
\label{fig:dyastimar}
\vspace{-0.6cm}
\end{figure}

The new software application DYASTIMA-R, which constitutes an extension of DYASTIMA uses the output provided by DYASTIMA, in order to calculate the energy that is deposited on the phantom and moreover the equivalent dose. Monte Carlo simulations are made in order to describe the particle interactions and the transport of the primary and secondary radiation through matter, especially through simulated media, such as the human body (phantom) and the aircraft shielding (optional), (Fig. \ref{fig:dyastimar}).

The absorbed dose D is calculated by using the mean energy dE deposited in a volume of mass dm at each step along a particle’s trajectory (eq. \ref{eq1}) while the equivalent dose H is calculated by using the absorbed dose D averaged over the phantom, multiplied by a quality factor related to the biological effectiveness of the radiation, $w_R$ (eq. \ref{eq2})\cite{cite1}, \cite{cite13}. Thus factor is defined as a function of the unrestricted linear energy transfer (LET) in water, which is the energy lost by a charged particle divided by the path length. Values of $w_R$ for different particles are given in Table \ref{tab1}.
\begin{table*}[ht!]
\begin{center}
\begin{tabular}{|c|c|}
\hline \textbf{Radiation Type} & \textbf{$w_R$} \\
\hline Photons & 1 \\
\hline Electrons and Muons & 1  \\
\hline Protons and charged Pions & 2\\
\hline Alpha particles, Fission fragments, Heavy Ions &  20\\
\hline Neutrons, $E_n\thinspace < \thinspace 1\thinspace MeV$ & $2.5+18.2 Exp(-(Log(E_n))^2/6)$\\
\hline Neutrons, $1\thinspace MeV\thinspace\leq\thinspace E_n \leq 50MeV$ & $5+17 Exp(-(Log(2E_n))^2/6)$\\
\hline Neutrons, $50\thinspace MeV\thinspace<\thinspace E_n$ &$2.5+3.25 Exp(-(Log(0.04E_n))^2/6)$\\
\hline
\end{tabular}
\caption{Radiation weighting factors $w_R$ for different particles \cite{cite1}.}
\label{tab1}
\end{center}
 \vspace*{-0.7cm}
\end{table*}
\begin{equation}\label{eq1}
D=\frac{d\overline{\epsilon}}{dm}
\end{equation}
\begin{equation}\label{eq2}
H_{T,R}=w_R D_{T,R}
\end{equation}
Since the radiation exposure field consists of different particles and energies, the total absorbed dose and total equivalent dose are calculated as the sum of the individual absorbed doses and equivalent doses respectively \cite{cite1}. 
\begin{equation}
D_T=\sum_R D_{T,R}
\end{equation}
\begin{equation}
H_T=\sum_R w_RD_{T,R}
\end{equation}
The GCR primary spectrum was extracted from CREME2009 for solar quiet conditions at solar maximum and minimum, which represent ambient conditions in the absence of solar energetic particle events \cite{cite14} \cite{cite15} \cite{cite16} \cite{cite17}. The magnetic threshold rigidity is of order of 0 GV. A cylindrical phantom (1.75 m height, 0.25 m radius) consisted of water is used. An optional airplane shell will be available soon, in order to study various shielding materials.

The application is under development and some preliminary results are presented. 

\section{FIRST RESULTS}

The study of the radiation exposure of aviators and passengers due to the contribution of GCR and SPEs is of great importance. For this reason, the scientific community was led to develop space weather forecasting and monitoring centers. The Athens Neutron Monitor Station (A.Ne.Mo.S.), participates as an expert group to European Space Agency (ESA SSA Space Radiation Center) providing timely and accurate warning for GLEs (GLE Alert). 

\begin{figure*}[h]
    \centering
    \includegraphics[width=\textwidth,bb=0 0 850 300]{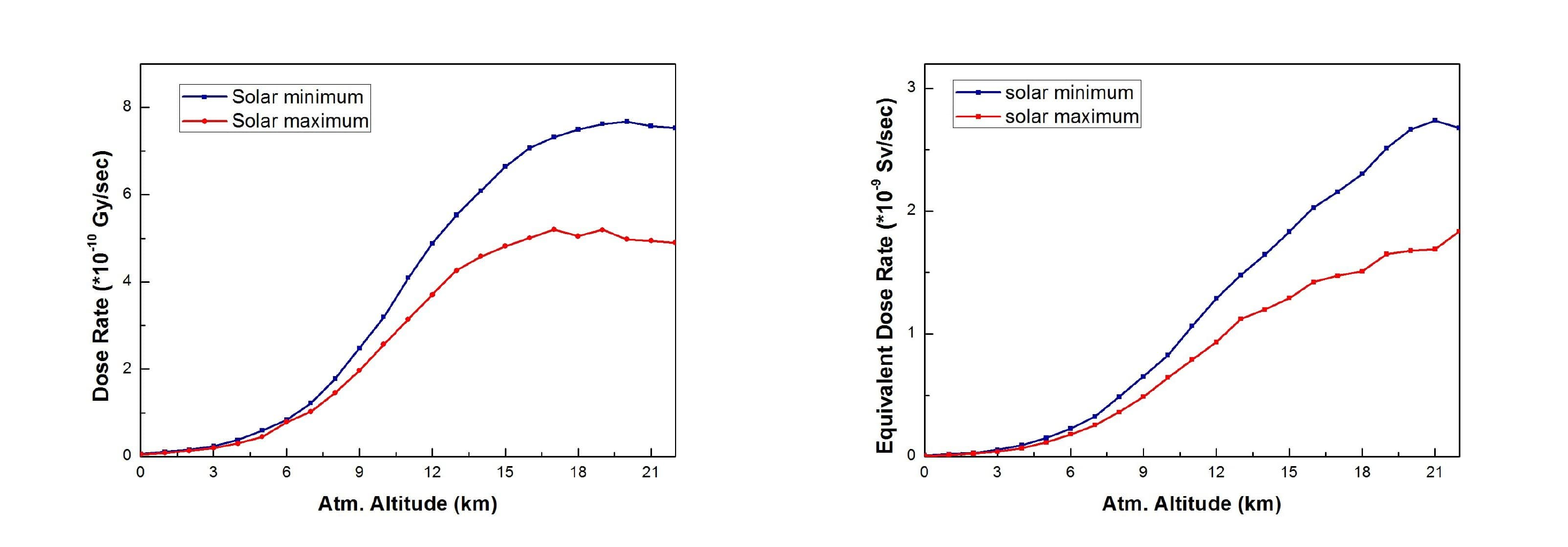}
    \vspace*{-0.6cm}
    \caption{Dose rate (left panel) and equivalent dose rate (right panel) for different atmospheric altitudes during solar minimum and solar maximum conditions.}\label{fig:dose}
\end{figure*}

\begin{figure*}[h]
    \centering
    \includegraphics[width=\textwidth,bb=0 0 850 300]{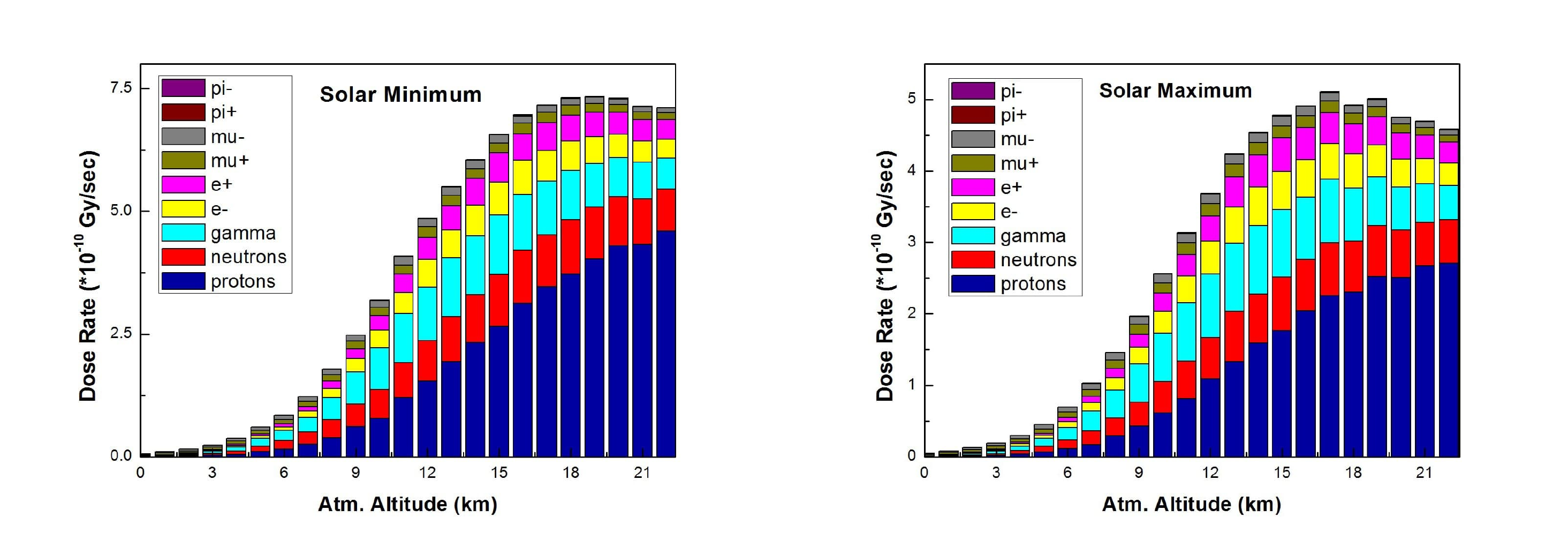}
    \vspace*{-0.6cm}
    \caption{Contribution of different radiation types to the total dose rate during solar minimum (left panel) and solar maximum conditions (right panel).}\label{fig:dose_types}
\end{figure*}

\begin{figure*}[h]
    \centering
    \includegraphics[width=\textwidth,bb=0 0 850 300]{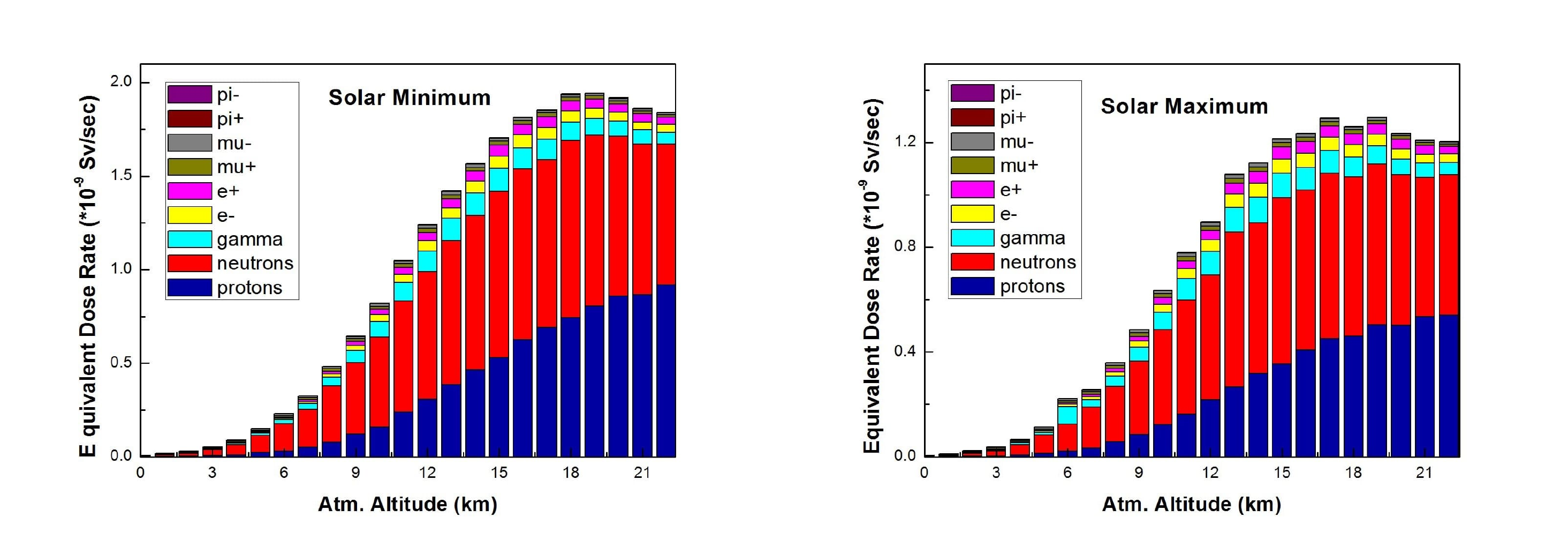}
    \vspace*{-0.6cm}
    \caption{Contribution of different radiation types to the total equivalent dose rate during solar minimum (left panel) and solar maximum conditions (right panel).}\label{fig:eq_dose_types}
\end{figure*}

In this work, the dose and equivalent dose are studied during the maximum and descending phase of solar cycle 23 and the ascending phase of solar cycle 24 (Fig. \ref{fig:dose}). Furthermore, the contribution of different radiation particles in total dose (Fig. \ref{fig:dose_types}) and total equivalent dose (Fig. \ref{fig:eq_dose_types}) is also studied. The main results of this study can be summarized as follows:

-The dose levels are directly related to the GCR particle intensities with the 11-year sunspot cycle and the 22-year solar magnetic cycle (Fig. \ref{fig:dose}). Since the GCR intensity is anti-correlated with the solar activity, the GCR exposure peaks at solar minimum and is lowest at solar maximum conditions \cite{cite13}. 

-The main contribution in the total dose is due to protons (Fig. \ref{fig:dose_types}), while the main contribution in total equivalent dose is due to neutrons (Fig. \ref{fig:eq_dose_types}).

For further studies, DYASTIMA-R will be applied during intense solar activity periods, such as SPE and Ground Level Enhancements (GLE) increase the energy deposit and the equivalent dose \cite{cite18}.

\section{Capabilities and Perspectives}
As DYASTIMA-R calculates the equivalent dose for various types of particles in different atmospheric altitudes and takes into account the phases of solar activity, as well as the geometry and shielding materials of the aircrafts, allowing the study of various flight scenarios. Therefore, it can be of great interest for:
\begin{itemize}
\vspace{-0.2cm}
\item Air-craft crews (pilots, flight attendants)
\vspace{-0.3cm}
\item Passengers (frequent travelers, pregnant women, children)
\vspace{-0.3cm}
\item Airlines and Tour Operators
\vspace{-0.3cm}
\item Air-craft manufacturers
\vspace{-0.3cm}
\item Legislators and Civil Aviation
\vspace{-0.3cm}
\end{itemize}

DYASTIMA-R will be combined with the GLE Alert system operated in A.Ne.Mo.S and ESA Space Radiation Center and soon will be provided as a tool for an extensive study of the radiation exposure during aircraft flights and manned space missions.
\vspace{0.3cm}
\section{Acknowledgements}
\vspace{-0.2cm}
Special thanks to the ESA Space Situational Awareness Program P2-SWE-1 ‘‘Space Weather Exert Centers: Definition and Developement“. We acknowledge the NMDB database (www.nmdb.eu), founded under the European Union's FP7 program (contract no. 213007) for providing cosmic ray data. A.Ne.Mo.S is supported by the Special Research Account of Athens University (70/4/5803).


\end{document}